\documentclass[11pt]{article}
\usepackage[T1]{fontenc}
\usepackage[utf8]{inputenc}
\usepackage{ae}
\usepackage{amsmath}

\usepackage{aecompl}
\usepackage{lmodern}
\usepackage{amsmath}
\usepackage{amssymb}
\usepackage{graphicx}
\title{Dynamical Structure of the Fields in the Light Cone Coordinates}
\author{K. Kargar$^{1}$, A. Shirzad $^{2,3}$ and M. Monemzadeh $^{1}$ \\
\small$^1$Department of Physics, University of Kashan, Kashan, Iran\\
 \small$^2$Department of Physics, Isfahan University of Technology, \small Isfahan, Iran\\ \small$^3$School of Particles and Accelerators, \\
 \small Institute for Research in Fundamental Sciences (IPM),
\small Tehran, Iran\\
}
\date{}
\begin{document}
\makeatletter
\newcommand*{\rom}[1]{\expandafter\@slowromancap\romannumeral #1@}
\maketitle
%
\begin{abstract}
It is well-known that additional constraints emerge in light cone
coordinates. We enumerate the number of physical modes in light
cone coordinates and compare it with conventional coordinates. We
show that the number of Schr\"odinger modes is divided by two in
light cone coordinates. We study the effect of this reduction in
the number of ladder operators acting on physical states of a
system. We analyse the scaler, spinor and vector field
theories carefully to see the effect of changes in the dynamical structure of these theories
from the view point of the reduction of Schr\"odinger modes in light-cone coordinates. In this way, we propose an alternative expansion of dynamical variables which defer from other literatures.
 
\end{abstract}
%
\newpage
\tableofcontents
\newpage
\section{Introduction}

Considering the various sub-groups of Poincar\' e group, in a
pioneer paper, Dirac in 1949\cite{dirac}, introduced three forms
for relativistic dynamics, instant form (I.F.), front form (F.F.)
and point form (P.F.). These forms are related to the various
choices of the time axis. The instant form is the usual choice of
the coordinate $x^0$ as the time coordinate, while in front form
$(x^0+x^3)/\sqrt{2}$ is chosen as the time coordinate. The front
form has special features with so many applications in theoretical
physics, specially in non perturbative $ QCD $\cite{brodeski},
string theory\cite{green}, gravity \cite{FOG} and so on. In the literature of high
energy physics, the front form is recognized with different names
such as ``$ Light\, Front $", ``$Infinite\, Momentum\, Frame $" and ``$
Light\, Cone $". In this paper we use light-cone. For a brief
review of light-cone quantization and its application in high
energy physics see Ref. \cite{heinzl}.

In the light-cone formulation of physical systems, the hyperplane ${x^ + } = {{({x^0} + {x^3})} \mathord{\left/
 {\vphantom {{({x^0} + {x^3})} {\sqrt 2 }}} \right.
 \kern-\nulldelimiterspace} {\sqrt 2 }}$ acts as the equal time hyperplane. The
light-cone coordinates are
${x^ \pm } = {{({x^0} \pm {x^3})} \mathord{\left/
 {\vphantom {{({x^0} \pm {x^3})} {\sqrt 2 }}} \right.
 \kern-\nulldelimiterspace} {\sqrt 2 }} = {{({x_0} \mp {x_3})} \mathord{\left/
 {\vphantom {{({x_0} \mp {x_3})} {\sqrt 2 }}} \right.
 \kern-\nulldelimiterspace} {\sqrt 2 }} = {x_ \mp }$ and ${x^ \bot } \equiv ({x^1},{x^2})$.\\
For an arbitrary four-vector $ A^\mu $ with components
$({A^0},\textbf{A})$ we
define the light-cone components as $({A^ + },{A^i},{A^ - }) = ({A^ +
},\tilde A)$ where $i=1,2$ and ${A^ \pm } = ({A^0} \pm {A^3})/\sqrt 2 $ and $\tilde A = \left( {{A^i},{A^ - }} \right)$. So,
for invariant space-time length element in Minkowski space we have
\begin{equation}
d{s^2} = dx_0^2 - dx_3^2 - dx_i^2 = 2 d{x_ + }d{x_ - } - dx_i^2,
\label{metric}
\end{equation}
which shows that the metric has non-diagonal elements.

Historically light-cone coordinates is well-known for particle physicists since it is used to derive some $QCD$ sum rules \cite{Fubini}\cite{Gell-Mann}. The large variety of applications of light-cone coordinates, come
from the advantage of relativistic dynamics of physical systems on
the hyperplane of the ${x_0} + {x_3} = \textrm{const}$. Dirac
mentioned some of these advantages. First, in light-cone coordinates the number of kinematical Poincar\' e generators are seven while in the conventional
formulation only six
are kinematical. Second, the non-diagonal form of light-cone
metric, enables us to separate the total energy of a system of
relativistic particles into center of mass energy and relative
energy \cite{heinzl}. This is different from the instant form, in
the sense that the appearance of the square root in the relation
of energy, ${P_0} = {({{\vec P}^2} + {M^2})^{1/2}}$, prohibits a
similar separation of variables. These advantages and specially
the latter one, have made the light-cone coordinates an appropriate
tool for calculating quantities such as wave functions.

One special feature of using light-cone coordinates is emergence of additional constraints compared to the conventional coordinates. We call these additional constraints \textit{light-cone constraints}. This change in constraint structure of the theory is well-known\cite{brodeski}. However, the number of light-cone constraints for a generic theory is not well understood yet. Physically we expect no change in the dynamical content of the theory upon changing the coordinates of space-time. So one needs to identify clearly the role of light-cone coordinates on the dynamical behaviour of the system. These are the main task of this work.

We will show explicitly that the light-cone constraints sit in place of half of the physical degrees of freedom. Hence, the number of dynamical degrees of freedom is divided by two, compared to the conventional coordinates. Although this phenomenon is met by physicists working on concrete models\cite{heinzl}, it is not clearly recognized as a general rule for an arbitrary model. We will show the light-cone constraints together with the remaining half of the dynamical equations of motion are equivalent to the whole equations of motion in conventional coordinates.

The next problem is how to choose the physical modes to be quantized in light-cone coordinates. For instance, some authors divide the momentum space into two parts and work with, say, the ${k_ - } > 0$ half of the momentum space\cite{kogut}. This happens when one insists on expanding the fields with the same combination as in conventional coordinates. In this paper we give another approach, in which we maintain the whole momentum space but put away half of the physical modes. In this approach, summation over spin in a spinor field and/or summation over polarization in a gauge field theory is no more necessary in light-cone coordinates. In other words, a light-cone observer is able to observe only one of the spin (polarization) states of a electron (photon).

In the reminder of this paper, we do the above task for the major type of physical theories which are quadratic or first order with respect to velocities. We show that in both types of theories the phenomenon of halving the number of dynamical modes is similar. In section 2, we find the general form of the constraint structure of a theory in light-cone coordinates and enumerate the number of dynamical variables.
We do this both for second order and first order Lagrangians. Section 3 denotes to the quantization procedure based on the symplectic approach of quantization, which is more or less a new approach in light-cone quantization. Sections 4 and 5 deal the same procedure for the special case of the spinor field theory and the vector field theory. In section 6 we discuss the case of Yang-Mills theories amd at the end of this section, we try embedding of non-Abelian Yang-Mills theories in light-cone coordinates using the BFFT method. The last section denotes out conclusions.

\section{ Number of dynamical variables}
As we mentioned earlier, formulation of theories in light cone
coordinates, leads to a different Hamiltonian structure in
comparison with conventional coordinates \cite{heinzl}. Since in light-cone coordinates ${x^ + }$ is the time coordinate,
the conjugate momentum is defined as
\begin{equation}
{\pi _{_{F.F.}}} = \frac{{\partial {\cal L}}}{{\partial ({\partial
_ + }\phi )}},
 \end{equation}
which differs from the ordinary instant form momentum
${\pi _{I.F}} = \frac{{\partial \mathcal{L}}}{{\partial ({\partial _0}\varphi )}}$ in the sense that
\begin{equation}
\pi_{FF}= \frac{1}{{\sqrt 2 }}({\pi _{I.F}} - {\partial _3}\varphi ).
 \end{equation}
In addition to different Hamiltonian structure, this point leads to a different number of dynamical variables.
We investigate the problem in turns for two major important field theoretic systems, i.e. quadratic 
Lagrangians and first order Lagrangians (with respect to the velocities).    
\subsection{Quadratic Lagrangian}
Consider a typical theory described by a set of dynamical fields $\phi_a (a =
1,2,...,n)$. Suppose the Lagrangian of
the theory is at most quadratic with respect to the partial
derivatives of the fields. Taking into account the Lorentz
invariance, the most general form of the kinetic term is
$g^{ab}{\partial _\mu }\phi_a \,{\partial ^\mu }\phi_b $ for some symmetric matrix $g$. In
conventional coordinates (Instant Form) we have $\mathcal{L} = 
g^{ab}(\partial _0 \phi_a  \partial _0 \phi_b - \nabla \phi_a . \nabla \phi_b)+\cdots $, and
definition of momenta (i.e. ${\pi^{^a} _{_{I.F}}} \equiv  2g^{ab}{\partial
_0}\phi_b $) gives no constraint for non singular $g$. In the light-cone
coordinates (Front Form), however, the kinetic term in the
Lagrangian is written as $ 2g^{ab}({\partial _ + }\phi_a \,{\partial _ -
}\phi_b - {\partial _ \bot }\phi_a .{\partial _ \bot }\phi_b) $, which gives the
conjugate momentum ${\pi^{^a} _{_{F.F}}} = 2g^{ab}{\partial _ - }\phi_b
$. Since there is no velocity in this relation we have the constraints 
\begin{equation}
\chi^a
\equiv{\pi^{^a} _{_{I.F}}} - 2g^{ab}{\partial _ - }\phi_b \simeq 0\,. \label{cons}
\end{equation}
Hence the non-diagonal form of the light-cone metric changes the constraint
structure of the system. If the original theory is not constrained
(i.g. Klein-Gordon theory), it will possess some new constraints,
while a system which is already a constrained system in
conventional coordinates (i.g. Electromagnetism) will
possess additional constraints due to the linearity of the
Lagrangian with respect to the velocities $\partial_+\phi_a$.

Suppose there are $ k $ first class and $ m $ second class constraints on the phase space 
in conventional coordinates. We also need $ k $
subsidiary conditions as gauge fixing conditions to reach the reduced phase space. Hence, there exist all together
$2k+m \equiv l$ conditions on the fields in phase space. The number of
degrees of freedom is therefore $2n-l$ in Hamiltonian formalism
and $n-l/2$ in Lagrangian formalism \cite{Henneaux}.

Now, by going to the light-cone coordinates the number of
remaining degrees of freedom in phase space should be divided by
2. The reason is as follows: the $n-l/2$ physical degrees of freedom
correspond to variables in the Lagrangian with truly quadratic terms
with respect to the velocities in the conventional coordinates. As we showed, the quadratic terms with respect to conventional velocities (i.e. ${({\partial _0}{\phi _a})^2}$) are replaced by terms ${\partial _ + }{\phi _a}{\partial _ - }{\phi _a}$ in light-cone coordinates which is linear with respect to velocities. 
Hence,
in the light-cone formulation of the
theory we will have $\frac{{2n - l}}{2}$ additional constraints which we call them "light cone constraints". 

The light cone constraints are second class in the
sense that their consistency with time determines the
corresponding Lagrange multipliers. Since each second class
constraint reduces one dynamical variable, we have
 \begin{equation}
{{\cal N}_{_C}^{^{F.F}}} = l+\frac{{2n - l}}{2} = n +
\frac{l}{2}.\label{heart}
 \end{equation}
where ${{\cal N}_{_C}^{^{F.F}}}$ is the
total number of constraints in front form. In this way half of the
dynamical variables of the phase space are omitted by the light
cone constraints and the number of degrees of freedom of the
theory reduces to $\frac{{2n - l}}{2}$.  In subsequent sections we
will see this effect for Klein-Gordon and electromagnetic field
theories.

However, note that we have not restricted the physical sector of
the theory in phase space by going to the light cone coordinates.
To see this we may try to project the additional constraints to
the conventional phase space to see if there is any possible
reduction. For this reason, we try to transform constraint ${\pi
_{_{F.F}}} - {\partial _ - }\phi  = 0$ from light-cone
coordinates, to conventional coordinates. By using chain rule for
${\pi _{F.F}}$ we obtain
 \begin{equation}
{\pi _{_{F.F}}} = \frac{{\partial {\cal L}}}{{\partial ({\partial
_0}\phi )}}\frac{{\partial ({\partial _0}\phi )}}{{\partial
({\partial _ + }\phi )}} + \frac{{\partial {\cal L}}}{{\partial
({\partial _3}\phi )}}\frac{{\partial ({\partial _3}\phi
)}}{{\partial ({\partial _ + }\phi )}} = \frac{1}{{\sqrt 2
}}({\partial _0}\phi  - {\partial _3}\phi ).
 \end{equation}
The right hand side of this equation is the same as ${\partial _ -
}\phi$ and so we will get the trivial relation $0=0$. So any
attempt to find an equivalent hyper-plane for the constraint
surface due to the light-cone constraints will lead to trivial
equations in the conventional coordinates. In other words, there is no
hyper plane in the conventional coordinate phase space equivalent
to the hyper plane of light-cone constraints. Therefore the change
in constraint structure in light-cone coordinates does not mean
that the classical phase space in conventional coordinates is
reduced.
\subsection{First Order Lagrangian} 

The most well known Lagrangian containing a Lorentz invariant first order  dynamical term includes 
${{\bar \psi }_\alpha }\gamma _{\alpha \beta }^\mu {\partial _\mu }{\psi _\beta }\,\,(\alpha ,\beta  = 1,2,...,n)$ 
as appears in the familiar Dirac Lagrangian. To study the constraint structure 
of such field theories, we investigate the symplectic matrix of this theory in 
conventional coordinates as well as light-cone coordinates. By considering 
${{\bar \psi }_\alpha }$ and ${\psi _\beta }$ as the independent variables 
of phase space, in  conventional coordinates the dynamical term ${{\bar \psi 
}_\alpha }\gamma _{\alpha \beta }^0{\partial _0}{\psi _\beta }$ gives the 
symplectic matrix as (see Appendix A)
\begin{equation}
\omega  = \left( {\begin{array}{*{20}{c}}
  0&{{{\gamma ^0}_{n \times n}}} \\ 
  { - {{\gamma ^0}_{n \times n}}}&0 
\end{array}} \right).
\end{equation}
Since $\det ({\gamma ^0}) \ne 0$, so ${\gamma ^0}$ does not have any null eigen-vector. In conventional 
coordinates, the number of  phase space degrees of freedom is $ 2n $. For example in ordinary Dirac fields,  
$ n=4 $ and the number of phase space variables is $ 8 $.

In light-cone coordinates we set ${\gamma ^\mu }{\partial _\mu } = {\gamma ^ + }{\partial _ + } +
 {\gamma ^ - }{\partial _ - } + {\gamma ^ \bot }{\partial _ \bot }$ where
\begin{equation}
{\gamma ^ \pm } = \frac{1}{{\sqrt 2 }}\left( {{\gamma ^0} \pm {\gamma ^3}} \right),
\end{equation}
The dynamical terms in the Lagrangian is ${{\bar \psi }_\alpha }\gamma _{\alpha
 \beta }^ + {\partial _+}{\psi _\beta }$ which gives the symplectic matrix as
\begin{equation}
\omega  = \left( {\begin{array}{*{20}{c}}
  0&{{{\gamma ^ + }_{n \times n}}} \\ 
  { - {{\gamma ^ + }_{n \times n}}}&0 
\end{array}} \right).
\end{equation}
In 4 dimensional space-time, all representations of Dirac matrices are unitarily equivalent, so it is sufficient 
to consider a specific representation and investigate the rank of ${\gamma ^ + }$. By choosing the 
chiral representation, we have \cite{peskin},
\begin{equation}
{\gamma ^ + }  = \frac{1}{{\sqrt 2 }}\left( {\begin{array}{*{20}{c}}
  0&0&2&0 \\ 
  0&0&0&0 \\ 
  0&0&0&0 \\ 
  0&2&0&0 
\end{array}} \right).
\end{equation}
which shows that ${\gamma ^ + }$ has two null eigen-vectors. Hence, in light-cone coordinates the 
number of degrees of freedom in the phase space  is $ n $ instead of $ 2n $. In the case of 4 dimensional Dirac field it is 4 instead of 8.

Thus, similar to the case of second order Lagrangian, the number of degrees of freedom is divided by two in the light-cone formulation of field theories with first order Lagrangian. Therefore, the number of Sch\" odinger modes in light-cone coordinates are half of those in the conventional coordinates.

\section{Quantization procedure} 

Let us see the effect of change in the constraint structure on the
classical dynamics, as well as the quantization procedure of the system. Classicaly we want to know what happens to
half of the degrees of freedom which are absent in the light cone coordinates.
In conventional coordinates we need to solve $2n -l$ first order
differential equations for the dynamical variables in phase space.
However, in light cone coordinates we have $n-l/2$ constraints
together with $n-l/2$ first order differential equations with
respect to time. Hence, the total number of equations at hand are
the same in both formalisms and the physical results are the same,
as it should be.

In fact, just the superficial features of the dynamical equations are different in two approaches; i.e. in conventional coordinates all
$2n-l$ equations of motion include derivatives with respect to
$x^0$, while in light cone coordinates $n-l/2$ constraints do not
include derivatives with respect to $x^+$ and the remaining
$n-l/2$ do include derivatives with respect to $x^+$.

Our next desire is finding suitable basis for the variables of the
reduced phase space in order to follow the dynamics of the
system. Suppose we are able to find a suitable basis for the whole
phase space of the system, in which imposing the constraints leads
to omitting a number of redundant variables. Such a basis is
recognized in the literature of constrained systems as the
Darbeaux basis \cite{Henneaux}. Hence, in comparison with conventional
coordinates, the additional light cone constraints lead to a
smaller reduced phase space with $n-l/2$ dynamical ($x^+$
dependent) modes. Then we should solve the equations of motion to
find the time dependence of physical modes in terms of
 $n-l/2$ independent Schr\"odinger modes (see Appendix A).

In fact, in a Darbeaux basis, in light-cone coordinates, the
procedure of solving the dynamics of the system will break into two
steps. First, imposing the light-cone second class constraints to
omit half of the degrees of freedom; and second, solving the
remaining equations of motion to find the Schr\"odinger modes.
Finally one needs to write the expansion of the fields in terms of
the Schr\"odinger modes.

Since the Schr\"odinger modes play the role of creation and
annihilation operators in the quantum theory, one may wonder
if different number of Schr\"odinger modes in light-cone
coordinates lead to a different quantum space of physical states.
We expect the physical quantities should not depend on the choice
of coordinate basis.

To answer this question we should consider the commutation
relations of ladder operators with the Hamiltonian of the system and
compare the results in light-cone and conventional coordinates. 
As a familiar example, we quantize the Klein-Gordon theory in
light-cone coordinates using the symplectic method. As we will see, in the light-cone coordinates, according
to Eq. \eqref{heart} we expect one constraint on the classical phase
space of the theory which affect the quantization procedure.

Klein-Gordon theory is introduced by the Lagrangian density
\begin{equation}
{\cal L} = \frac{1}{2}({\partial _\mu }\varphi \,{\partial ^\mu }\varphi  - {m^2}{\varphi ^2}) = {\partial _ + }\varphi \,{\partial _ - }\varphi  - \frac{1}{2}{({\partial _ \bot }\varphi )^2} - \frac{1}{2}{m^2}{\varphi ^2}.
\end{equation}
The conjugate light-cone momentum is
\begin{equation}
\pi  \equiv \frac{{\partial {\cal L}}}{{\partial ({\partial _ + }\varphi )}} = {\partial _ - }\varphi \label{KGcon},
\end{equation}
which introduces the primary constraint $\chi  \equiv \pi  -
{\partial _ - }\varphi \simeq 0$  on the phase space. The total
Hamiltonian \cite{dirac2} reads
 \begin{equation}
{H_T} = \int {{d^3}{\tilde x}\,\left( {\frac{1}{2}  {{({\partial _
\bot }\varphi )}^2} + \frac{1}{2}{m^2}{\varphi ^2} + u(x)\chi (x)}
\right)},
 \end{equation}
where ${u(x)}$ is the Lagrange multiplier. Assume the equal time fundamental Poisson brackets as
 \begin{align}
&{\left\{ {\varphi ({\tilde x}),\varphi ({\tilde x}')}
\right\}_{{x^ + }}} = {\left\{ {\pi ({\tilde x}),\pi
({\tilde x}')} \right\}_{{x^ + }}} = 0, \nonumber \\
&{\left\{ {\varphi ({\tilde x}),\pi ({\tilde x}')} \right\}_{{x^ + }}}
 = \delta ({x^ - } - {{x'}^ - }){\delta ^2}({x^ \bot } - {{x'}^ \bot }).
 \end{align}
Since the constraint $\chi $ considered at different points
constitute a system of second class constraints, the consistency
condition $ {\partial _ + }\chi (x) = {\{ \chi (x),{H_T}\} _{{x^ +
}}} = 0 $ will not give a secondary constraint; instead, it
determines the Lagrange multiplier via the equation
\begin{equation}
2{\partial _ - }u(x) = ({\partial _ \bot }{\partial _ \bot }\varphi  - {m^2}\varphi ).
\end{equation} 
To impose the single constraint $\chi(x)$ on the fields, it is more suitable to use the
following Fourier expansions
 \begin{align}
\varphi  = \frac{1}{{{{(2\pi )}^{3/2}}}}\int {{d^3}
{\tilde k}\,a({\tilde k},{x^ + })\,{e^{i{\tilde k}.
{\tilde x}}}}, \nonumber \\
\pi  = \frac{1}{{{{(2\pi )}^{3/2}}}}\int {{d^3}{\tilde k}\,c
({\tilde k},{x^ + })\,{e^{ - i{\tilde k}.{\tilde x}}}},\label{KGexp}
 \end{align}
 The physical modes are $\,a({\tilde k},{x^ + })$ and
$c({\tilde k},{x^ + })$. Imposing light-cone constraint $ \pi  -
{\partial _ - }\varphi  = 0 $ on the expansions \eqref{KGexp} gives
\begin{equation}
c({\tilde k},{x^ + }) =  - i{k_ - }a( - {\tilde k},{x^ + }).\label{s-1}
\end{equation}
In contrast with conventional coordinates where  there are two
physical modes, in light-cone coordinates we have only one
independent physical mode which we assume to be $a(\tilde k,x^+)$. Eq. \eqref{s-1} shows that 
$c({\tilde k})$ is determined in terms of $a({\tilde k})$. Now
using Eq. \eqref{two form}, in the appendix A, to construct symplectic two-form, we have
\begin{equation}
\Omega  = \int {{d^3}{\tilde k}\,\left( { - 2i{k_ - }} \right)\,da( - {\tilde k} , {x^ + }) \wedge da({\tilde k} , {x^ + })}.
\end{equation}
Hence, the Dirac brackets of the physical modes are
\begin{equation}
{\left\{ {a({\tilde k},{x^ + }),a({\tilde k}',{x^ + })} \right\}_{D.B}} = \frac{{ - 1}}{{2i{k_ - }}}{\delta ^3}({\tilde k} + {\tilde k}').\label{KGDB}
\end{equation}
In terms of the physical modes $a(\tilde k,x^+)$, the canonical
Hamiltonian is
\begin{equation}
H_c = \int {{d^3}\tilde k\frac{1}{2}( k_ \bot ^2 +
{m^2})\,a ( \tilde k,{x^ + })\,a( - \tilde k,{x^ + })}.
\end{equation}
Using the canonical Hamiltonian, we are able to write
equations of motion of the physical modes as
\begin{equation}
\dot a({\tilde k},{x^ + }) = \left\{ {a,H} \right\} =
i{\omega _ + }a({\tilde k},{x^ + })\label{KG EOM modes},
\end{equation}
where
\begin{equation}
 {\omega _ + }  \equiv  \frac{{k_ \bot ^2 + {m^2}}}{{2{k_ - }}} \label{on shell condition} .
\end{equation}
The solution of Eq. \eqref{KG EOM modes} is
\begin{equation}
a({\tilde k},{x^ + }) = a({\tilde k},0)\,{e^{i{\omega _ + }{x^ + }}}.
\end{equation}
In contrast with the conventional coordinates where we deal
with two coupled first order differential equations of motion, in light-cone coordinates we have only one
differential equation. As we mentioned earlier, imposing
the light-cone constraint \eqref{KGcon}, is equivalent to solving
one equation of motion of ordinary coordinates. The original
fields can be expanded in terms of the Schr\"odinger modes $ a({\tilde
k},0) $ as
\begin{align}
&\varphi ({\tilde x},{x^ + }) = \frac{1}{{{{(2\pi )}^{3/2}}}}\int {{d^3}{\tilde k}\,a({\tilde k},0)\,{e^{ikx}}}, \nonumber \\
&\pi ({\tilde x},{x^ + }) = \frac{1}{{{{(2\pi )}^{3/2}}}}\int {{d^3}{\tilde
k}\,( - i{k_ - })a( - {\tilde k},0)\,{e^{ - ikx}}}. \label{KGexp final}
\end{align}

By using Eq. \eqref{KGDB} and Dirac quantization prescription $\left\{
{\,\,\,\,,\,\,\,\,} \right\}\,\, \to \,\, - i\left[
{\,\,\,\,,\,\,\,\,} \right]$, the quantum commutators of the
Schr\"odinger modes as well as original fields can be written as
\begin{align}
&\left[ {a({\tilde k}),a({\tilde k}')} \right] = \frac{{ - 1}}{{2{k_ - }}}{\delta ^3}({\tilde k} + {\tilde k}') \label{KGcr},\\
&\left[ {\varphi ({\tilde x},{x^ + }),\pi ({\tilde y},{x^ + })} \right] = \frac{1}{2}\,{\delta ^3}({\tilde x} - {\tilde y}), \\
&\left[ {\varphi ({\tilde x},{x^ + }),\varphi ({\tilde y},{x^ + })} \right] = \frac{1}{2}\,\theta ({x^ - } - {y^ - }){\delta ^2}({x^ \bot } - {y^ \bot }).\label{KGCCR}
\end{align}
where $\theta ({x^ - } - {y^ - })$ is the Heaviside step function.

In contrast to conventional coordinates, Eq. \eqref{KGCCR} shows
that the field $ \phi $ does not commute with itself on equal
light-cone time hyper plane. Let us investigate this property carefully. We want to find the
commutation relation \eqref{KGCCR} from the non-equal time
commutation relation of Klein-Gordon fields in conventional
coordinates. In conventional coordinates we have\cite{peskin}
\begin{equation}
{\left[ {\varphi (x),\varphi (y)} \right]_{I.F}} =
\int {{d^3}\textbf{k}\,\frac{1}{{2{\omega _\textbf{k}}}}\left( { - {e^{ - ik(x - y)}} + {e^{ik(x - y)}}} \right)},
\end{equation}
which can be written covariantly as
\begin{equation}
{\left[ {\varphi (x),\varphi (y} \right]_{I.F}} = \int {{d^4}k\,\delta ({k^2} - {m^2})\,\theta ({k_0})} \left( { - {e^{ - ik(x - y)}} + {e^{ik(x - y)}}} \right).\label{IFnet}
\end{equation}
Note that the subscription I.F in Eq. \eqref{IFnet} is no more necessary in covariant form of the commutation relations. Hence, we can transform this integral to light-cone coordinate. By transforming $\delta ({k^2} - {m^2})$ to light-cone coordinates and integrating over ${{k_ + }}$
 we have
\begin{equation*}
{\left[ {\varphi (x),\varphi (y)} \right]} = \int {d{k_ - }{d^2}{k_ \bot }\,\frac{1}{{2{k_ - }}}\left( { - {e^{ - ik(x - y)}} + {e^{ik(x - y)}}} \right)\theta \left( {\frac{{{k_ + } + {k_\_}}}{{\sqrt 2 }}} \right)\left| \begin{array}{l}
\\
 {k_ + } = \frac{{k_ \bot ^2 + {m^2}}}{{2{k_ - }}}
\end{array} \right.}\ \ .
\end{equation*}
Putting ${x^ + } = {y^ + }$, we can find the equal light-cone time commutation relations as
\begin{equation}
\left[ {\varphi ({\tilde x},{x^ + }),\varphi ({\tilde y},{x^ + })}
\right] = \frac{1}{2}\,\theta ({x^ - } - {y^ - }){\delta ^2}({x^
\bot } - {y^ \bot }),
\end{equation}
which is exactly the commutation relation \eqref{KGCCR} we obtained
by direct calculation in the light-cone coordinates. This simple
result which shows consistency of formulation of Klein-Gordon
theory in light-cone and conventional coordinate systems, although
expected intuitively, is not shown explicitly in the literature
yet. Note that the transformation from light-cone to conventional
coordinates is not an ordinary Lorentz transformation.

Now we will turn back to the problem of interpreting different
number of ladder operators in light-cone and conventional coordinates. Let see how we can
interpret different number of Schr\"odinger modes in light-cone
coordinates?

Consider the commutation relation of
ladder operators with the Hamiltonian in both coordinates. In conventional coordinates we have 
\begin{align}
&\left[ {A(\textbf{k}),{A^\dag }(\textbf{k})} \right] = {(2\pi )^3}{\delta ^3}(\textbf{k} - \textbf{k}'), \\
&\left[ {H,A(\textbf{k})} \right] =  - {\omega _\textbf{k}}A(\textbf{k}), \label{1}\\
&\left[ {H,{A^\dag }(\textbf{k})} \right] = {\omega
_\textbf{k}}{A^\dag }(\textbf{k}).\label{2} \end{align} 
where
${\omega _\textbf{k}}$ is the time component of the momentum 4-vector. As we see,
the sign of the right hand sides of Eqs. \eqref{1} and \eqref{2} are different for the annihilation and creation operators.
In conventional coordinates, the on shell condition reads ${\omega _\textbf{k} ^2} = k_0^2 = {\textbf{k}^2} +
{m^2}$; hence, the sign of the spacial components of momentum do not determine the sign of $\omega
_\textbf{k} $.

In the light-cone coordinates, however, the number of the ladder
operators is divided by 2 and the commutation relations are
\begin{align}
&\left[ {a({\tilde k}),a({\tilde k}')} \right] = \frac{1}{{2{k_ - }}}{\delta ^3}({\tilde k} + {\tilde k}'), \nonumber \\
&\left[ {H,a({\tilde k})} \right] =  - {\omega _{\tilde k}}a({\tilde k}).\label{ccr}
\end{align}
Remembering Eq. \eqref{on shell condition}, shows that the sign of $ {\omega
_{\tilde k}} $ depends on the sign of ${k_ - }$. This property
divides the momentum space into two parts ${k_ - } > 0$ and
${k_ - } < 0$ where $a({\tilde k})$ is a creation operator in ${k_ - } > 0$ region and an
annihilation operator in ${k_ - } < 0$ region.
This point of view differs from conventional approach \cite{srivastava, burkardt, popov} which insist on introducing two sets
of ladder operators for creation and annihilation on the price of restricting the physical domain
of momentum coordinate $k_-$ to the region $k_->0$.

Let us investigate the effect of light-cone ladder operators on
the total momentum of the system. Using the definition of the energy-momentum
tensor, $T_\nu ^\mu  = \frac{{\partial \mathcal{L}}}{{\partial
({\partial _\mu }\varphi )}}{\partial _\nu }\varphi  -
\mathcal{L}\delta _\nu ^\mu $, the components of momentum in the light-cone coordinates are
\begin{align}
\begin{gathered}
  {P^ + } = \int {{d^3}\tilde x\,\pi \,{\partial _ - }\varphi }  = \int {{d^3}\tilde k\,(k_ - ^2)\,a( - k)\,a(k)},  \hfill \\
  {P^i} = \int {{d^3}\tilde x( - \,\pi \,{\partial _ - }\varphi )}  =- \int {{d^3}\tilde k\, {k_ - }{k_i}\,a( - k)\,a(k)}, \,\,\,\,\,\,i = 1,2 \hfill \\
\end{gathered}
\end{align}
Using commutation relation \eqref{KGcr} we have
\begin{align}
&\left[ {{P^ + },\,a({\tilde k})} \right] = \left[ {{P_ - },\,a({\tilde k})} \right] = {k_ - }a({\tilde k}).\\
&\left[ {{P_i},\,a({\tilde k})} \right] =  - \left[ {{P^i},\,a({\tilde k})} \right] = {k_i}a({\tilde k}).
\end{align}
These relations verifies the interpretation of $a({\tilde k}) $ with $ {k_ - } > 0 $ ($ {k_ - } < 0 $) as  
creation (annihilation) operators.

\section{Symplectic light-cone Quantization of Spinor fields}
Dirac theory is introduced by the first order Lagrangian density
\begin{equation}
\mathcal{L} = \bar \psi (i{\gamma ^\mu }{\partial _\mu } - m)\psi. \label{Dirac Lagrangian}
\end{equation}
To quantize this theory in light-cone coordinates, it is convenient to use a decomposition of spinor space by the projection operators\cite{kogut},
\begin{equation}
{\Lambda ^ \pm } = \frac{1}{2}{\gamma ^ \mp }{\gamma ^ \pm } = \frac{1}{{\sqrt 2 }}{\gamma ^0}{\gamma ^ \pm },
\end{equation}
which project the spinor field $ \psi $ to $ {\psi _ \pm } = {\Lambda ^ \pm }\psi $. Using the identities,
\begin{equation}
{\gamma ^0}{\gamma ^ + } = {\gamma ^ - }{\gamma ^0},\,\,\,\,\,{\Lambda ^ \pm }{\Lambda ^ \mp } = 0,
\end{equation}
we have $ {\gamma ^0}{\psi _ + } = \frac{{\sqrt 2 }}{2}{\gamma ^ + }{\psi _ + } $.
Hence, the Lagrangian density of the Dirac field decomposes as,
\begin{equation}
\mathcal{L} = i\sqrt 2 \psi _ + ^\dag {\partial _ + }{\psi _ + } + i\sqrt 2 \psi _ - ^\dag {\partial _ - }{\psi _ - } - \psi _ - ^\dag (m + i{\gamma ^i}{\partial _i}){\gamma ^0}{\psi _ + } - \psi _ + ^\dag (m + i{\gamma ^i}{\partial _i}){\gamma ^0}{\psi _ - }.
\end{equation}
In this way, the Lagrangian \eqref{Dirac Lagrangian} can be written as
\begin{equation}
\mathcal{L} = i\sqrt 2 \psi _ + ^\dag {\partial _ + }{\psi _ + } - \psi _ - ^\dag {\chi _1} - {\mathcal{H}_c}
\end{equation}
where the density of canonical Hamiltonian is,
\begin{equation}
{\mathcal{H}_c} = \frac{1}{{\sqrt 2 }}\psi _ + ^\dag (m + i{\gamma ^i}{\partial _i}){\gamma ^ - }{\psi _ - }.\label{dirac hamiltonian}
\end{equation}
In the above Lagrangian, the only dynamical variables are ${\psi _ + }$ and $\psi _ + ^\dag $, while the equations of motion for the the variables $\psi _ - ^\dag $ and ${\psi _ - }$ give the constraints,
\begin{align}
\begin{gathered}
  {\chi _1} \equiv i{\partial _ - }{\psi _ - } - \frac{1}{2}(m + i{\gamma ^i}{\partial _i}){\gamma ^ + }{\psi _ + } \approx 0, \hfill \\
  {\chi _2} \equiv i{\partial _ - }\psi _ - ^\dag  + \frac{1}{2}(m\psi _ + ^\dag  - i{\partial _i}\psi _ + ^\dag {\gamma ^i}){\gamma ^ - } \approx 0. \hfill \\ \label{DiracConstraints}
\end{gathered} 
\end{align}
In order to write a suitable mode expansion of the fields ${\psi _ +
}$ and $\psi _ + ^\dag $ we look for a complete set of
eigenfunctions of the Hamiltonian of the first quantized
theory.
In conventional coordinates, $u(\textbf{k}){e^{\textbf{ik.x}}}$
and $v(\textbf{k}){e^{-\textbf{ik.x}}}$ are the eigenfunctions
of Dirac Hamiltonian ${h_D}$,
\begin{equation}
{h_D} =  - i{\gamma ^0}{\gamma ^i}{\partial _i} + m{\gamma ^0}\,\,\,\,\,\,\,\,\,i = 1,2,3\,,
\end{equation}
 with the energy eigenvalues ${E_k}$
and $ - {E_k}$ respectively\cite{peskin}. Actually, the solutions of the eigenvalue equations ${h_D}\psi (x) =  \pm {E_k}\psi (x)$ can be considered as $u(k){e^{-ik.x}}$
and $v(k){e^{ik.x}}$ such as,
\begin{equation}
\left( {{\gamma ^\mu }{k_\mu } - m} \right)u(k) = 0 \label{u},
\end{equation}
\begin{equation}
\left( {{\gamma ^\mu }{k_\mu } + m} \right)v(k) = 0 \label{v}.
\end{equation}
Each of the equations \eqref{u} and \eqref{v} have two independent solutions distinguished by the eigenvalues of the component of spin operator, say in the third direction, i.e. ${\Sigma ^3}$. Hence, for every solution of \eqref{u} and \eqref{v} we can decompose ${u^1}$ and ${u^2}$ as well as ${v^1}$ and ${v^2}$ by using the projection operators 
\begin{equation}
{\cal{S}^ \mp } = \frac{1}{2}\left( {1 \pm {\Sigma ^3}} \right)\label{spin projection}
\end{equation}
In this way for the Dirac fields $ \psi (x)$ and $\bar \psi (x)$ with 8 independent phase space variables, we can set the 8 eigenspinors $\left\{ {{u_s},{v_s},u_s^\dag ,v_s^\dag } \right\}$ for $ s=1,2 $.
So , the summation over spin indices is necessary in the conventional coordinates.

On the other hand, in the light-cone coordinates, due to additional constraints \eqref{DiracConstraints}, 
the dimension of the reduced phase space is 4. So, in order to expand independent phase space variables in term of energy eigenfunctions, we need 4 energy eigenfunctions of the Hamiltonian operator. To do this, notice that the Dirac light-cone Hamiltonian operator can be recognized from the canonical Hamiltonian \eqref{dirac hamiltonian} as
\begin{equation}
{h^{L.C}_D} \equiv \frac{1}{{\sqrt 2 }}(m + i{\gamma ^i}{\partial _i}){\gamma ^ - }.
\end{equation}
Using the plane wave solutions \eqref{u} and \eqref{v} we can introduce ${u_ \pm }(k) = {\Lambda ^ \pm }u(k)$ and ${v_ \pm }(k) = {\Lambda ^ \pm }v(k)$. Then it is easy to see,
\begin{align}
\begin{gathered}
  {h^{L.C}_D}u_ + ^{}(k) = {k_ + }u_ + ^{}(k), \hfill \\
  {h^{L.C}_D}u_ - ^{}(k) = 0, \hfill \\
  {h^{L.C}_D}v_ + ^{}(k) =  - {k_ + }v_ + ^{}(k), \hfill \\
  {h^{L.C}_D}v_ - ^{}(k) = 0. \hfill \\
\end{gathered}
\end{align} 
 As is seen, spinors $u_ - ^{}(k)$ and $v_ - ^{}(k)$ are ruled out from the eigenspinors of $ {h^{L.C}_D} $. On the other hand spinors $\left\{ {{u_ + },{v_ + },u_ + ^\dag ,v_ + ^\dag } \right\}$ form a basis for the four dimensional space of variables ${\psi _ + }$ and $\psi _ + ^\dag $. In other words, in light-cone coordinates in contrast with conventional coordinates, there is the natural projection operator $ {\Lambda ^ \pm } $ for the energy eigenspinors.\\ 
 In this way there is no need to use the spin projection operators \eqref{spin projection} to distinguish the degenerate spinors. Hence, there is no spin summation in expansion of the Dirac fields. Now we are able to expand the dynamical fields ${\psi _ + }$ and
 $\psi _ + ^\dag $ in the basis $ {u_ + }(k){e^{ - i\tilde{k}.\tilde{x} }} $ and $ {v_ + }(k){e^{i\tilde{k}.\tilde{x}}} $ and their conjugates as follows,
\begin{equation}
{\psi _ + }(\tilde{x},{x^ + }) = \int {{d^3}\tilde{k}\left( {A(\tilde{k},{x^ + }){u_ + }(k){e^{ - i\tilde{k}.\tilde{x}}} + {B^\dag }(\tilde{k},{x^ + }){v_ + }(k){e^{i\tilde{k}.\tilde{x}}}} \right)},
\end{equation}\label{psi_+}
\begin{equation}
\psi _ + ^\dag (\tilde{x},{x^ + }) = \int {{d^3}\tilde{k}\left( {{A^\dag }(\tilde{k},{x^ + })u_ + ^\dag (k){e^{i\tilde{k}.\tilde{x}}} + B(\tilde{k},{x^ + })v_ + ^\dag (k){e^{ - i\tilde{k}.\tilde{x}}}} \right)} \label{psi_+^dag}.
\end{equation}
Before going through the expansion of the fields, let us see what has happened to the state of the eigenstates with spinors $ {u_ + }(k) $ and $ {v_ + }(k) $. For this reason consider the spin states of energy eigenfunctions in the rest frame ${k_r} = \frac{1}{{\sqrt 2 }}\left( {m,0,0,m} \right)$. By choosing the rest frame in the relations \eqref{u} and \eqref{v} we simply have the solutions,
\begin{equation}
{u_ + } = \left( \begin{gathered}
  1 \hfill \\
  0 \hfill \\
  0 \hfill \\
  1 \hfill \\
\end{gathered}  \right)\,\,\,\,\,\,\,\,\,\,,\,\,\,\,\,\,\,\,\,\,{v_ + } = \left( \begin{gathered}
  1 \hfill \\
  0 \hfill \\
  0 \hfill \\
   - 1 \hfill \\
\end{gathered}  \right)\,.
\end{equation}
Compare these with the conventional basis ${u^1}$ and ${u^2}$ as ${v^1}$ and ${v^2}$ in the rest frame as,
\begin{equation}
{u^1} = \left( \begin{gathered}
  1 \hfill \\
  0 \hfill \\
  1 \hfill \\
  0 \hfill \\
\end{gathered}  \right)\,\,\,\,,\,\,\,\,{u^2} = \left( \begin{gathered}
  0 \hfill \\
  1 \hfill \\
  0 \hfill \\
  1 \hfill \\
\end{gathered}  \right)\,\,\,\,,\,\,\,\,{v^1} = \left( \begin{gathered}
  1 \hfill \\
  0 \hfill \\
   - 1 \hfill \\
  0 \hfill \\
\end{gathered}  \right)\,\,\,\,,\,\,\,\,{v^2} = \left( \begin{gathered}
  0 \hfill \\
  1 \hfill \\
  0 \hfill \\
   - 1 \hfill \\
\end{gathered}  \right)
\end{equation}
where ${\Sigma ^3}{u^1} = {u^1}$, ${\Sigma ^3}{u^2} =  - {u^2}$, ${\Sigma ^3}{v^1} = {v^1}$ and ${\Sigma ^3}{v^2} =  - {v^2}$. It is easy to see ${\Lambda ^ + }{u^1} = {\Lambda ^ + }{v^1}$ and ${\Lambda ^ + }{u^2} =  - {\Lambda ^ + }{v^2}$. This says that the spin states for positive and negative frequency solutions of conventional coordinates are no longer independent after projecting with the operator ${\Lambda ^ + }$. Hence, we can recognize the combination of spin states as,
\begin{equation}
\begin{gathered}
  {u_ + } = {\Lambda ^ + }({u^1} + {u^2}) = {\Lambda ^ + }({v^1} - {v^2}), \hfill \\
  {v_ + } = {\Lambda ^ + }({v^1} + {v^2}) = {\Lambda ^ + }({u^1} - {u^2}). \hfill \\
\end{gathered}
\end{equation}
So, energy eigenfunctions ${u_ + }$ and ${v_ + }$ are projections of some combinations of spin states. Therefore, Schr\"odinger modes (or equivalently ladder operators in quantum theory) create and annihilate particles and antiparticles in specific superposition of spin states. This property is in contrast to the quantized Dirac fields in conventional coordinates.

For the dependent fields ${\psi _ - }$ and $\psi _ - ^\dag $, using the constraints \eqref{DiracConstraints} we simply have,
\begin{equation}
{\psi _{-} }(\tilde{x},{x^{+} }) = \int d^3\tilde{k}\left( A(\tilde{k},x^+ )u_ - (k)e^{ - i\tilde{k}.\tilde{x}} + B^\dag (\tilde{k},{x^ + })v_ - (k)e^{i\tilde{k}.\tilde{x}},\right)
\end{equation}
\begin{equation}
\psi _ - ^\dag (\tilde{x},{x^ + }) = \int {{d^3}\tilde{k}\left( {{A^\dag }(\tilde{k}.\tilde{x},{x^ + })u_ - ^\dag (k){e^{i\tilde{k}.\tilde{x}}} + B(\tilde{k},{x^ + })v_ - ^\dag (k){e^{ - i\tilde{k}.\tilde{x}}}} \right)},
\end{equation}
where
\begin{align}
\begin{gathered}
  {u_ - }(k) = \frac{{m + {\gamma ^i}{k_i}}}{{2{k_ - }}}{\gamma ^ + }{u_ + }(k), \hfill \\
  {v_ - }(k) = \frac{{m - {\gamma ^i}{k_i}}}{{2{k_ - }}}{\gamma ^ + }{v_ + }(k). \hfill \label{u-v} \\
\end{gathered}
\end{align}
These relations are also in consistency with the relations \eqref{u} and \eqref{v}.\\
To find out the odd Poisson brackets of physical modes, we construct the symplectic two-form $ \Omega  = \int {{d^3}x\,i\sqrt 2 d} \psi _ + ^\dag  \wedge d{\psi _ + } $ by using Eqs. \eqref{psi_+} and \eqref{psi_+^dag} as follows
\begin{equation}
 \Omega = \int {{d^3}\tilde{k}\,i\sqrt 2 \left( {d{A^\dag }(\tilde{k},{x^ + }) \wedge dA(\tilde{k},{x^ + }) - d{B^\dag }(\tilde{k},{x^ + }) \wedge dB(\tilde{k},{x^ + })} \right)}.
\end{equation}
So the odd Poisson brackets of physical modes read
\begin{align}
\begin{gathered}
  {\left\{ {A(\tilde{k},{x^ + }),{A^\dag }(\tilde{k}',{x^ + })} \right\}_ + } = \frac{{ - i}}{{\sqrt 2 }}{\delta ^3}(\tilde{k} - \tilde{k}'), \hfill \\
  {\left\{ {B(\tilde{k},{x^ + }),{B^\dag }(\tilde{k}',{x^ + })} \right\}_ + } = \frac{i}{{\sqrt 2 }}{\delta ^3}(\tilde{k} - \tilde{k}')\,. \hfill \\ \label{DiracDB}
\end{gathered}
\end{align}
The canonical Hamiltonian \eqref{dirac hamiltonian} in terms of physical modes can be written as,
\begin{equation}
{\mathcal{H}_c} = \int {{d^3}\tilde{k}\frac{{{m^2} + k_i^2}}{{2{k_ - }}}\left( {{A^\dag }(\tilde{k},{x^ + })A(\tilde{k},{x^ + }) - {B^\dag }(\tilde{k},{x^ + })B(\tilde{k},{x^ + })} \right)}.
\end{equation}
Using the above Hamiltonian and the algebra \eqref{DiracDB}, the equations of motion of physical modes become
\begin{align}
\begin{gathered}
  {\partial _ + }A(\tilde{k},{x^ + }) =  - i{\omega _ + }A(\tilde{k},{x^ + }), \hfill \\
  {\partial _ + }{A^\dag }(\tilde{k},{x^ + }) = i{\omega _ + }{A^\dag }(\tilde{k},{x^ + }), \hfill \\
  {\partial _ + }B(\tilde{k},{x^ + }) =  - i{\omega _ + }B(\tilde{k},{x^ + }), \hfill \\
  {\partial _ + }{B^\dag }(\tilde{k},{x^ + }) = i{\omega _ + }{B^\dag }(\tilde{k},{x^ + }). \hfill \label{DiracEOMs} \\
\end{gathered}
\end{align}
where ${\omega _ + } = \frac{{{m^2} + k_i^2}}{{2{k_ - }}}$. By writing the solutions of Eqs. \eqref{DiracEOMs} in terms of Schr\"odinger modes and inserting them into the Eqs. \eqref{psi_+} and \eqref{psi_+^dag} we have 
\begin{align}
\begin{gathered}
  {\psi _ + }(\tilde{x},{x^ + }) = \int {{d^3}\tilde{k}\left( {A(\tilde{k}){u_ + }(k){e^{ - ikx}} + {B^\dag }(\tilde{k}){v_ + }(k){e^{ikx}}} \right)},  \hfill \\
  \psi _ + ^\dag (\tilde{x},{x^ + }) = \int {{d^3}\tilde{k}\left( {{A^\dag }(\tilde{k})u_ + ^\dag (k){e^{ikx}} + B(\tilde{k})v_ + ^\dag (k){e^{ - ikx}}} \right)}.  \hfill \\
\end{gathered}
\end{align}
Similar results can be written for ${\psi _ - }$ and $\psi _ - ^\dag $ where $u_-$ and $v_-$ are derived as in Eqs. \eqref{u-v}.\\ 
Using $\psi  = {\psi _ + } + {\psi _ - }$ and $u = {u_ + } + {u_ - }$, the expansions of original Dirac fields becomes,
\begin{align}
\begin{gathered}
  {\psi _{}}(\tilde{x},{x^ + }) = \int {{d^3}\tilde{k}\left( {A(\tilde{k})u(k){e^{ - ikx}} + {B^\dag }(\tilde{k})v(k){e^{ikx}}} \right)},  \hfill \\
  \psi _{}^\dag (\tilde{x},{x^ + }) = \int {{d^3}\tilde{k}\left( {{A^\dag }(\tilde{k})u_{}^\dag (k){e^{ikx}} + B(\tilde{k})v_{}^\dag (\tilde{k}){e^{ - ikx}}} \right)}.  \hfill \\ \label{Dirac exp}
\end{gathered}
\end{align}
As we mentioned earlier, in light-cone coordinates, the summation over spin states is no longer necessary in the expansions of the fields.
 This property is due to additional constraints \eqref{DiracConstraints} which appears in light-cone coordinates. Also same situation arises in light-cone electromagnetic theory where we need not too choose any polarization vector to quantize this theory.

\section{Symplectic light-cone Quantization of Vector fields}

The familiar electromagnetic theory is a gauge theory with two first class constraints the in
conventional coordinates. Let us investigate the constraint
structure of this theory in light-cone coordinates.

The Lagrangian $- \frac{1}{4}{F^{\mu \nu }}{F_{\mu \nu }}$ of
electromagnetic theory should be written in light-cone coordinates
as
 \begin{equation}
 {\cal L} = \frac{1}{2}{F_{ +  - }}{F_{ +  - }} +
  {F_{ - i}}{F_{ + i}} - \frac{1}{4}{({F_{ij}})^2}.
 \end{equation}
The conjugate momenta are
\begin{equation}
{\pi ^ + } = \frac{{\partial {\cal L}}}{{\partial ({\partial _ + }{A_ + })}} = 0,\label{EMmom1}
\end{equation}
 \begin{equation}
 {\pi ^ - } = \frac{{\partial {\cal L}}}{{\partial ({\partial _ + }{A_ - })}} = {F_{ +  - }},\label{EMmom2}
 \end{equation}
 \begin{equation}
 {\pi ^i} = \frac{{\partial {\cal L}}}{{\partial ({\partial _ +
 }{A_ - })}} = {F_{ - i}}\ \ \ \ \ i = 1,2.\label{EMmom3}
 \end{equation}
which give the primary constraints in the light-cone phase space as
follows
\begin{equation}
{\chi _0} \equiv {\pi ^ + } \simeq 0,
\end{equation}
 \begin{equation}
 {\chi _i} \equiv {\pi ^i} - {F_{ - i}} \simeq 0\,\,\,\,\,\,i =
 1,2.
 \end{equation}
The total Hamiltonian reads
 \begin{equation}
{H_T} = \int {{d^3}{\tilde{x}}\left( {\frac{1}{2}{{({\pi ^ - })}^2} + {\pi ^
- }{\partial _ - }{A_ + } + {\pi ^i}{\partial _i}{A_ + } +
\frac{1}{4}{F_{ij}}{F_{ij}} + u(x){\pi ^ + } + {v_i}(x)({\pi ^i} -
{F_{ - i}})} \right)},
 \end{equation}
where $u(x)$ and $v_i(x)$ are Lagrange multipliers. Assuming the
fundamental Poisson brackets as
 \begin{equation}
\left\{ {{A_\mu }(\tilde{x},{x^ + }),{\pi ^\nu }(\tilde{y},{x^ + })} \right\} =
\delta _\mu ^\nu \,{\delta ^3}(\tilde{x} - \tilde{y}),
 \end{equation}
consistency condition of the constraint $\chi_0$ gives the
secondary constraint
 \begin{equation}
{\phi _0}\equiv {\partial _i}{\pi ^i} + {\partial _ - }{\pi ^ - }
\approx 0,
 \end{equation}
while consistency of the constraints $\chi_i$ determines the
Lagrange multipliers $v_i$  via
 \begin{equation}
2{\partial _ - }{v_i}= {\partial _i}{\pi ^ - } -{\partial
_j}{F_{ij}}.
 \end{equation}
Consistency of the secondary constraint $\phi$ does not lead to a
new constraint. Hence we have two first class constraints ${\chi
_0}$ and ${\phi _0}$ and two second class constraints ${\chi _i}$.
Comparing with our general discussion on the number of degrees of
freedom in section 2, here we have $n=4$ physical fields $A^\mu$
with $k=2$ first class constraints $\pi_0$ and $\partial_i\pi_i (i=1,2,3)$, and
no second class constraint in the conventional coordinates. The
first class constraints $\chi_0$ and $\phi_0$ above are similar to
the first class constraints in conventional coordinates. The
number of phase space degrees of freedom in conventional
coordinates is $2n-2k=4$. However,  the number of degrees
of freedom is divided by two in light cone coordinates due to two
additional constraints $\chi_i$ which has not any counterpart in
conventional coordinates.

To construct the reduced phase space of the system we need two
gauge fixing conditions conjugate to our two first class
constraints. We begin with the gauge fixing condition $\omega_1
\equiv {A_ - } \approx 0$, which is, in fact, conjugate to the secondary constraint 
$\phi _0$ . The consistency condition of this gauge, i.e. $\partial _ + A_ -\approx 0$, gives
 \begin{equation}
\omega_2 \equiv \pi^-  + \partial_-A_ + \approx 0
 \end{equation}
which is the second required gauge fixing condition.  Consistency of $\omega_2$ gives an
equation to determine the Lagrange multiplyer $u(x)$. By imposing
the 4 constraints and 2 gauge fixing conditions one obtains a
reduced phase space with only two field variables. To determine the smallest set of independent
physical modes, we should write a suitable expansion of fields and
conjugate momenta and impose these constraint on them. As usual, the Fourier expansion is the suitable one, i.e.
 \begin{equation}
{A_\mu (x) } = \frac{1}{{{{(2\pi )}^{3/2}}}}\int
{{d^3}\tilde{k}\,{e^{i\tilde{k}.\tilde{x}}}{a_\mu }(\tilde{k},{x^ + })},\label{s-4}
  \end{equation}
 \begin{equation}
 {\pi ^\mu  (x)} = \frac{1}{{{{(2\pi )}^{3/2}}}}\int
 {{d^3}\tilde{k}\,{e^{ - i\tilde{k}.\tilde{x}}}{b^\mu }(\tilde{k},{x^ + })}.\label{s-5}
 \end{equation}
Imposing the constraints and guage fixing conditions on the physical modes $a_\mu$ and $b_\mu$ we
find the following six conditions, 
\begin{equation}
 \left\{ \begin{array}{l}
{b^ + }(\tilde{k},{x^ + }) = 0,\\
{b^i}(\tilde{k},{x^ + }) =  - i\left( {{k_ - }{a_i}( - \tilde{k},{x^ + }) - {k_i}{a_ - }( - \tilde{k},{x^ + })} \right),\\
 - i{k_ - }{b^ - }(\tilde{k},{x^ + }) = i{k_i}{b^i}(\tilde{k},{x^ + }),\\
{a_ - }(\tilde{k},{x^ + }) = 0,\\
{b^ - }(\tilde{k},{x^ + }) = i{k_ - }{a_ + }( - \tilde{k},{x^ + }).
\end{array} \right.
 \end{equation}
There remain two independent physical modes which can be
chosen as ${a_1}(\tilde{k},{x^ + })$ and ${a_2}(\tilde{k},{x^ + })$. Here, noticing that the field elements are real functions, we construct a linear superposition of these independent modes in a conjugate way as,
\begin{equation}
a(k,{x^ + }) = {a_1}(k,{x^ + }) + i{a_2}(k,{x^ + })
\end{equation}
\begin{equation}
{a^\dag }(k,{x^ + }) = {a_1}( - k,{x^ + }) - i{a_2}( - k,{x^ + })
\end{equation}
Rewriting physical modes according to this set of independent modes, we have,
\begin{align}
\begin{gathered}
  {a_ - }(k,{x^ + }) = {b^ + }(k,{x^ + }) = 0 \hfill \\
  {a_ + }(k,{x^ + }) =  - (\frac{{{k_1} - i{k_2}}}{{2{k_ - }}})a(k,{x^ + }) - (\frac{{{k_1} + i{k_2}}}{{2{k_ - }}}){a^\dag }( - k,{x^ + }) \hfill \\
  {b^1}(k,{x^ + }) = \frac{{ - i{k_ - }}}{2}\left( {a( - k,{x^ + }) + {a^\dag }(k,{x^ + })} \right) \hfill \\
  {b^2}(k,{x^ + }) = \frac{{ - i{k_ - }}}{2}\left( {a( - k,{x^ + }) - {a^\dag }(k,{x^ + })} \right) \hfill \\
  {b^ - }(k,{x^ + }) = \frac{i}{2}\left( {({k_1} - i{k_2})a( - k,{x^ + }) + ({k_1} + i{k_2}){a^\dag }( - k,{x^ + })} \right) \hfill \\ 
\end{gathered}
\end{align} 
To this end
we can construct the symplectic two form as
 \begin{equation}
\Omega  = \int {{d^3}\tilde{x}\,2\,(d{\pi ^\mu} \wedge d{A_\mu})}  =\int {{d^3}k\,( - i{k_ - })\left( {d{a^\dag }(k,{x^ + }) \wedge da(k,{x^ + })} \right)} 
 \end{equation}
Using the inverse of symplectic matrix (see the appendix) we find
the Dirac brackets of physical modes as
 \begin{equation}
\left\{ {a(k,{x^ + }),{a^\dag }(k',{x^ + })} \right\} = \frac{{ - 1}}{{i{k_ - }}}\delta ({k_ - } - {{k'}_ - })\,{\delta ^2}({k_ \bot } - {{k'}_ \bot })\, .\label{EMDB}
 \end{equation}
The canonical Hamiltonian in terms of the physical modes can be written as,
 \begin{equation}
{H_c} = \int {{d^3}k(\frac{{k_1^2 + k_2^2}}{2})\left( {{a^\dag }(k,{x^ + })a(k,{x^ + })} \right)}. \label{QED-Ham}
 \end{equation}
In contrast to conventional coordinates \cite{Henneaux}, the Hamiltonian  
\eqref{QED-Ham} is diagonal in terms of the transverse modes
$a(\tilde{k},{x^ + })$ and ${a^\dag} (\tilde{k},{x^ + })$. In other words, the transverse modes
appear in light cone coordinates in a natural way and we need not
to choose any polarization direction to quantize the theory. In
fact, by eliminating the redundant modes due to the light cone
constraints we need not to assume any polarization direction (as is done for instance in light-cone spinor field where  the summation over spin indices has been eliminated).

Using the Hamiltonian of Eq. \eqref{QED-Ham} and the Dirac brackets
\eqref{EMDB}, the equations of motion of physical modes read
  \begin{equation}
{\partial _ + }{a}(\tilde{k},{x^ + }) = \left\{ {{a},{H_c}} \right\} = i{\omega
_ + }{a}(\tilde{k},{x^ + }),\label{s-2}
  \end{equation}
 \begin{equation}
{\partial _ + }{a^\dag}(\tilde{k},{x^ + }) = \left\{ {{a^\dag},{H_c}} \right\} = -i{\omega
_ + }{a^\dag}(\tilde{k},{x^ + }),\label{s-3}
 \end{equation}
where ${\omega _ + } = \frac{{k_1^2 + k_2^2}}{{2{k_ - }}}$. Inserting the
solutions of Eqs. \eqref{s-2} and \eqref{s-3} 
in the expansions \eqref{s-4} and \eqref{s-5} of fields, we find non vanishing components of electromagnetic fields as,
 \begin{equation}
{A_ +(x) } = \frac{1}{{{{(2\pi )}^{3/2}}}}\int {{d^3}k\, {(-1)} {e^{ikx}}\left( {(\frac{{{k_1} + i{k_2}}}{{2{k_ - }}})a(k) - (\frac{{{k_1} + i{k_2}}}{{2{k_ - }}}){a^\dag }( - k)} \right)}, \label{Em exp 1}
 \end{equation}
 \begin{equation}
{A_1(x)} = \frac{1}{{{{(2\pi )}^{3/2}}}}\int {{d^3}k\,{e^{ikx}}\frac{1}{2}\left( {a(k) + {a^\dag }( - k)} \right)}, \label{Em exp 2}
 \end{equation} 
 \begin{equation}
{A_2(x)} = \frac{1}{{{{(2\pi )}^{3/2}}}}\int {{d^3}k\,{e^{ikx}}\frac{{ - i}}{2}\left( {a(k) - {a^\dag }( - k)} \right)}. \label{Em exp 3}
 \end{equation}
where $a(k)\equiv a_i(k,0)$ are Schr\"odinger modes. For non vanishing components of momentum fields we have also the following expansions
 \begin{equation}
{\pi ^ - } = \frac{1}{{{{(2\pi )}^{3/2}}}}\int {{d^3}k\,{e^{ - ikx}}\,\frac{i}{2}\left( {({k_1} - i{k_2})a( - k) + ({k_1} + i{k_2}){a^\dag }( - k)} \right)} 
 \end{equation}
  \begin{equation}
{\pi ^1} = \frac{1}{{{{(2\pi )}^{3/2}}}}\int {{d^3}k\,{e^{ - ikx}}\frac{{ - i{k_ - }}}{2}\left( {a( - k) + {a^\dag }(k)} \right)} 
 \end{equation}
  \begin{equation}
 {\pi ^2} = \frac{1}{{{{(2\pi )}^{3/2}}}}\int {{d^3}k\,{e^{ - ikx}}\frac{{ - i{k_ - }}}{2}\left( {a( - k) - {a^\dag }(k)} \right)} 
 \end{equation}
 Using  brackets \eqref{EMDB} we are able to calculate the Dirac
brackets of the fields and conjugate momenta which is in complete agreement with known results\cite{srivastava}. These relations can be seen in Appendix B.

\section{Light-Cone Quantization of Yang-Mills Theories}
In this section, we try to quantize non-Abelian Yang-Mills theories using symplectic method of quantization in light-cone coordinates. We will show that, we are not able to impose constraints on Fourier expansion of dynamical fields but this is not a light-cone quantization problem. We will try to embed light-cone non-Abelian Yang-Mills theories using the BFFT method of quantization\cite{19}.

Yang-Mills theories are theories for describing the behaviour of
elementary gauge particles intermediating the physical
interactions given by the Lagrangian density
\begin{equation}
{\cal L} =  - \frac{1}{4}F_a^{\mu \nu }F_{\mu \nu }^a,
\end{equation}
where $F_{\mu \nu }^a = {\partial _\mu }A_\nu ^a - {\partial _\nu
}A_\mu ^a + g\,f_{bc}^aA_\mu ^bA_\nu ^c$, in which $g$ is the
coupling constant, and $f_{ab}^c$ are the structure constants of a
Lie group (i.e. the gauge group). The index $a$ runs over
$1,2,...,N$ where $ N $ is the number of generators of the gauge
group.

In conventional coordinates, this theory includes $N$ first class
primary constraints and $N$ secondary first class
constraints. Taking into account $2N$ gauge fixing
conditions, there remain $8N-4N$ degrees of freedom\cite{Henneaux}.
However, in light cone coordinates according
to Eq.\eqref{heart}, we expect to have $2N$ additional second
class constraints which is half number of dynamical degrees of
freedom.

The Lagrangian density in light-cone coordinate reads
 \begin{equation}
{\cal L} = \frac{1}{2}F_{ +  - }^aF_{ + - }^a + F_{ - i}^aF_{ +
i}^a - \frac{1}{4}{(F_{ij}^a)^2}.
 \end{equation}
The conjugate momentums are similar to Eqs. \eqref{EMmom1} - \eqref{EMmom3} with the
additional subscript $a$ on the momentum fields $\pi_a^\mu$
conjugate to the fields $A^a_\mu$. Again
$\phi _a^0 \equiv \pi _a^ +  \simeq 0$ and $\phi _a^i \equiv \pi
_a^i - F_{ - i}^a \simeq 0$ are primary constraints and the total
Hamiltonian reads
 \begin{align} \begin{array}{l}
{H_T} = \int {{d^3}\tilde{x}(\frac{1}{2}{{(\pi _a^ - )}^2}+\pi _a^ -
 {{({D_ - })}^{ab}}A_ + ^b + \pi _a^i{{({D_i})}^{ab}}A_
 + ^b + \frac{1}{4}F_{ij}^aF_{ij}^a + } \\
\,\,\,\,\,\,\,\, + {u^d}(x)\pi _d^ +  + v_i^e(x)(\pi _e^i - F_{ -
i}^e)),
 \end{array}
\end{align}
where $u^d(x)$ and $v_i^e(x)$ are Lagrange multipliers and
${({D_\nu })^{ab}} \equiv \delta _b^a{\partial _\nu } -
g\,f_{bc}^aA_\nu ^c$ is the covariant derivative. Assuming the
fundamental Poisson brackets as
 \begin {equation} \left\{ {A_\mu ^a(\tilde{x},{x^ + }),\pi
_b^\nu (\tilde{y},{x^ + })} \right\} = \delta _\mu ^\nu \delta
_b^a\,{\delta ^3}(x - y),
 \end{equation}
the consistency condition  ${\partial _ + }\phi _a^0\approx 0$
gives a set of secondary first class constraints as
 \begin{equation}
{\chi ^a} \equiv {({D_i})^{ab}}\pi _b^i + {({D_ - })^{ab}}\pi _b^
-  \approx 0.
 \end{equation}
Consistency condition of this secondary constraints holds
identically. Consistency of the constraints $\phi _a^i$ determines
the Lagrange multipliers $v_i^e$ via the relations
 \begin{equation}
2{({D_ - })^{ab}}v_i^b + {({D_j})^{ab}}F_{ij}^b -
{({D_i})^{ab}}\pi _a^ - = 0
 \end{equation}
To construct the reduced phase space of the system we need $ 2N $ gauge fixing
conditions conjugate to our $ 2N $ first class constraints $\phi _a^0$ and ${\chi ^a}$. 
To choose required gauge fixing conditions we simply generalize the Electromagnetic gauge
 fixing conditions and choose $\omega _1^a \equiv A_ - ^a \approx 0$ conjugate to $\phi _a^0$.
 The consistency condition of $ \omega _1^a $ gives another gauge fixing condition as
\begin{equation}
\omega _2^a \equiv \pi _a^ -  + {\partial _ - }A_ + ^a \approx 0
\end{equation}
Hence,  there are altogether $ 6N $ conditions on the fields and conjugate momenta as
\begin{align}
\left\{ \begin{gathered}
  \phi _a^0 \equiv \pi _a^ +  \approx 0, \hfill \\
  \phi _a^i \equiv \pi _a^i - F_{ - i}^a \approx 0\,\,\,\,\,\,i = 1,2 \, , \hfill \\
  {\chi ^a} \equiv {({D_i})^{ab}}\pi _b^i + {({D_ - })^{ab}}\pi _b^ -  \approx 0, \hfill \\
  \omega _1^a \equiv A_ - ^a \approx 0, \hfill \\
  \omega _2^a \equiv \pi _a^ -  + {\partial _ - }A_ + ^a \approx 0. \hfill \\ 
\end{gathered}  \right.
\label{YMcons}
\end{align}

\subsection{Symplectic Method}
In the scaler theory and Electromagnetic field, we simply choose Fourier expansions of fields to find the independent physical modes. But in the non-Abelian Yang-Mills theories, there are some non-linear terms in constraints such as
\begin{equation}
\phi _a^i = \pi _a^i - F_{ - i}^a = \pi _a^i - {\partial _\mu }A_\nu ^a + {\partial _\nu }A_\mu ^a - g\,f_{bc}^aA_\mu ^bA_\nu ^c = 0
\end{equation} 
Due to the existence of non-linear terms such $g\,f_{bc}^aA_\mu ^bA_\nu ^c$ we are not able to impose this
constraint on the Fourier expansion of fields to construct the reduced phase space. However, this problem is not 
the problem of light-cone quantization and is the fundamental problem of quantization of non-Abelian Yang-Mills
theories. To quantize this theory, one can set the limit $g = 0$ and quantize the theory perturbatively but attempt
to quantize the theory directly fails due to lack of an appropriate expansion of fields which enables us impose
the constraints.\\
In the next subsection we try to embednon-Abelian Yang-Mills theories to an extended phase space to see the problem from another point of view.

\subsection{BFFT Method}
In this subsection, we try to embed non-Abelian Yang-mills theories in an extended phase space in which second class constraints, i.e. $\phi _a^i$
 become first class constraints using BFFT method\cite{19,abarghoo}.  In this method, first of all, we need to extend phase space by adding some extra fields $(q,p) \oplus \eta $. the number of these auxiliary fields are equal to the number of second classs constraints appear in original phase space.
  By introducing ${\omega _{\alpha \beta }}$ as the algebra of new variables $\{\eta _{\alpha} ,\eta _\beta\} = {\omega _{\alpha \beta }}$ and ${\Delta _{\alpha \beta }} = \left\{ {\tau _{\alpha} ^{(0)},\tau _{\beta} ^{(0)}} \right\}$ where ${\tau _{\alpha} }(q,p,\eta )$ is our constraint in embedded phase space and ${\tau _{\alpha} ^{(n)}}$ are the $n$th order of expansion of embedded constraints according to new variables $\eta$, we have,
 \begin{eqnarray}
\tau _\alpha ^{(1)} = \chi _\alpha ^\beta (q,p){\eta _\beta }\label{BFTexpand}, \\
{\Delta _{\alpha \beta }} + \chi _\alpha ^\gamma {\omega _{\gamma \lambda }}\chi _\beta ^\lambda  = 0.\label{BFTmaster}
 \end{eqnarray}
 We are able to choose $\eta$ in such a way that the second class constraints become first class in new phase space. To find out this first class constraints, we have to solve the master equation, Eq.\eqref{BFTmaster}, according to $\chi _\alpha ^\gamma$. We have,
 \begin{eqnarray}
\Delta _{\alpha \beta }^{ab} = \left[ {\begin{array}{*{20}{c}}
{ - {\alpha ^{ab}}}&0\\
0&{ - {\alpha ^{ab}}}
\end{array}} \right]{\delta ^3}(x - y),\\
{\alpha ^{ab}} = 2{({D_ - })^{ab}} = (2\delta _b^a{\partial _ - } - gf_{bc}^aA_ - ^c).
 \end{eqnarray}
  In order to solve Eq.\eqref{BFTmaster} we need to guess  ${\omega _{\gamma \lambda }}$. As shown in \cite{19} by choosing ${\omega _{\gamma \lambda }}$ as below, the BFFT method become finite order. From now on, for simplicity, we drop out gauge indices $a$ and $b$.\\
 \begin{equation}
{\omega _{\gamma \lambda }} = \left[ {\begin{array}{*{20}{c}}
1&{ - 1}\\
1&1
\end{array}} \right],\label{omega}
 \end{equation}
To solve \eqref{BFTmaster} we consider $\chi _\alpha ^\gamma$ as,
 \begin{equation}
 \chi _\alpha ^\gamma  = \left[ {\begin{array}{*{20}{c}}
{{a_1}}&{{a_2}}\\
{{a_3}}&{{a_4}}
\end{array}} \right].
 \end{equation}
By putting these relations in \eqref{BFTmaster}, we see that we have 4 unknown parameters $a_i$ and 3 equations, so we need to guess at least 1 of $a_i$s to solve equations. Actually, different guess for these parameters and different solutions transform to each other by canonical transformations. We have,
 \begin{equation}
 {a_1} = 0\,\,\,\,\,\& \,\,\,\,\,{a_4} = 0\,\,\,\,\,\, \Rightarrow \,\,\,\,\,{a_3} = {a_2} = \sqrt {{\alpha ^{ab}}}.
 \end{equation}
 So first class constraints in embedded phase space at first order can be written as,
 \begin{equation}
\left\{ \begin{array}{l}
\tau _1^{(1)} = {(2\delta _b^a{\partial _ - } - gf_{bc}^aA_ - ^c)^{1/2}}\eta _2^b.\\
\tau _2^{(1)} = {(2\delta _b^a{\partial _ - } - gf_{bc}^aA_ - ^c)^{1/2}}\eta _1^b.
\end{array}, \right.\label{BFFTconstraints}
 \end{equation}
 By choosing ${\omega _{\gamma \lambda }}$  as the Eq.\eqref{omega}, as shown in \cite{19}, the embedding become truncated and higher order in expansion ${\tau _\alpha }(q,p,\eta ) = \sum\limits_{n = 0}^\infty  {\tau _\alpha ^{(n)}} $ vanishes.\\
 In relations \eqref{BFFTconstraints}, we see the square root of operator ${({D_ - })^{ab}}$ which does not make sense well. To avoid this ambiguity we are able to expand perturbatively with the assumption 
 $\frac{{gf_{bc}^aA_ - ^c}}{{2\delta _b^a{\partial _ - }}} \ll 1$ which is a true assumption in $QCD$ for large momentums. Doing this and dropping out terms higher than $\frac{{gf_{bc}^aA_ - ^c}}{{2\delta _b^a{\partial _ - }}} \ll 1$ , we have,
 
 \begin{equation}
\left\{ \begin{gathered}
  \tau _1^{(1)} = \left( {{{(2\delta _b^a{\partial _ - })}^{1/2}} - (\frac{1}{2}\frac{{gf_{bc}^aA_ - ^c}}{{{{(2\delta _b^a{\partial _ - })}^{1/2}}}})} \right)\eta _2^b \hfill \\
  \tau _2^{(1)} = \left( {{{(2\delta _b^a{\partial _ - })}^{1/2}} - (\frac{1}{2}\frac{{gf_{bc}^aA_ - ^c}}{{{{(2\delta _b^a{\partial _ - })}^{1/2}}}})} \right)\eta _1^b \hfill \\ 
\end{gathered}  \right.
 \end{equation}
 
 The above relations are embeded constraints in extended phase space which are first class.Note that we obtained these constraint perturbatively and these are not exact.
 
 Comparing two methods discussed in this section, we see that in the case of symplectic method for quantizing non-Abelian Yang-Mills theories, due to non linear terms in constraints \eqref{YMcons}, there is not any appropriate expansion of fields which enables us to obtain the smallest set of physical modes unless we put $g=0$. But in the BFFT method, although we cannot find embeded constraint exactly, we can find them perturbatively. This is remarkable consequence.

\section{Conclusion}

The appearance of additional constraints on a field theory in light-cone coordinates is well-known. However, for the expansions of fields and conjugate momenta in light-cone coordinates according to physical modes, authors use the usual expansions of conventional coordinates with an extra Heaviside step function. This step function divides the momentum space into two parts which enables us to choose say $k_{-}>0$ part of the momentum space.
In this paper, we have proposed alternative expansions of the fields and conjugate momenta on the whole momentum space for the scaler, fermionic and vector fields. Our expansion is  based on the fact that the number of independent physical modes must be equal to the number of degrees of freedom in phase space.
To do this, we exactly have investigated the dynamical structure of phase space variables by enumerating the number of degrees of freedom as well as independent physical modes as follows.

First of all, we have shown that non-diagonal form of
the light-cone metric causes changes in the constraint structure of the
field theories described by the quadratic and first order Lagrangians. We showed exactly that half of the dynamical equations of motion are replaced by the
light-cone constraints, hence the number of dynamical degrees of freedom is divided by
two, comparing with the conventional coordinates. Although this phenomenon is met by
physicists working on concrete models[4], it is not clearly recognized as a general role
for an arbitrary model. We showed that the light-cone constraints together with
the remaining half of the dynamical equations of motion are equivalent to the whole
equations of motion in conventional coordinates.

Second, since the number of independent physical modes are equal to the number of degrees of freedom, using the symplectic method of quantization, we  chose the most appropriate set of independent physical modes to expand the phase space variables. By imposing the constraints on the phase space variables to obtain the reduced phase space, we priori have solved half of the equations of motion. By solving the remaining equations of motion, we obtained Schr\"odinger modes. Then we showed that each one of  Schr\"odinger modes can play the role of creation or annihilation operator depending on the sign of the $k_{-}$ component of the momentum vector. 

At the end, using symplectic method of quantization and analysing the dynamical structure of the phase space variables, we have proposed alternative expansions of the fields in the whole momentum space with true number of physical modes. Notice that the number of independent physical modes must be equal to the number of independent phase space variables.
In the case of scaler field, we have obtained relations \eqref{KGexp final} which shows that only one set of Schr\"odinger modes i.e. $a(k)$ act as ladder operators. In the cases of fermionic and vector fields, our expansions have more significant remarks.\\
In the case of fermionic fields, we have obtained relations \eqref{Dirac exp} for phase space variables. Additional light-cone constraints eliminate summation over spin indices in these expansions. We showed that ladder operators in the quantum theory create and annihilate particles and antiparticles in a specific superposition of spin indices.\\
Similar situation arises in the case of the vector field. As illustrated in the relations (\ref{Em exp 1},\ref{Em exp 2},\ref{Em exp 3}), the summation over polarization states is no longer necessary in the expansions of the fields and conjugate momenta.

We also investigated the constraint structure of the Non-Abelian Yang-Mills theories in light-cone coordinates; and showed that the number of degrees of freedom is again half of those of conventional coordinates. Due to the existence of non-linear terms such $g\,f_{bc}^aA_\mu ^bA_\nu ^c$ in the expressions of the constraints, we are not able to impose the
constraints on the Fourier expansion of the fields to construct the reduced phase space. However, this problem is not 
due to light-cone quantization and is the fundamental problem of quantization of non-Abelian Yang-Mills
theories. To quantize this theory, one can set the limit $g = 0$ and quantize the theory perturbatively.  
However, any attempt
to quantize the theory directly fails due to lack of appropriate expansions of the fields which enable us to impose
the constraints.

Same problem arose in BFFT embedding of non-Abelian Yang-Mills Theories due to square root of operator 
${(2\delta _b^a{\partial _ - } - gf_{bc}^aA_ - ^c)^{1/2}}$, so we are not able to quantize the theory exactly and non perturbatively.\\ 
\\
\\
\\

\textbf{Acknowledgement}\\
The authors thank H. Ghaemi for valuable discussions.

\newpage
\appendix
\section*{Appendix A}
\subsection*{Symplectic Method of Quantization}
In this appendix, we briefly review the symplectic method of
quantization which is proposed originally by Faddeev and Jackiw
\cite{faddeev} and is showed that is equivalent to Dirac`s method
of quantization of constrained systems\cite{mojiri}. For a
comprehensive review of this method see  Ref. \cite{bakhshi}.

For practical purpose, the symplectic method is based on the
following steps:
\begin{enumerate}
\item[A)]  First, we should determine the complete constraint structure of systems\cite{dirac2}. This means that we should determine the primary constraints by calculating conjugate momenta. Then by applying the consistency conditions and defining fundamental Poisson brackets of canonical variables, we should obtain the secondary constraints. In this step we also need to construct the canonical Hamiltonian of the system.\\
After investigating the constraint structure of the system, we should classify constraints as first and second class constraints using fundamental Poisson brackets. According to a conjecture by Dirac\cite{dirac2}, first class constraints are the generators of gauge transformations. In this way, we should fix the gauges by imposing additional gauge fixing conditions on the system. In Ref.\cite{shirzadgf}, the essential requirements for an appropriate gauge fixing conditions are given.

\item[B)] In the second step, we should propose appropriate expansions of the fields and which enables us to impose constraints and gauge fixing conditions to get the reduced phase space. This step determines physical modes as the smallest set of time dependent variables which uniquely describe every state of the classical system.
In many familiar cases the Fourier expansion is an appropriate choice.

\item[C)] This step is the most important part of the quantization procedure in which we find the canonical commutation relation of fields. The symplectic two form is defined as\cite{bakhshi}
\begin{equation}
\Omega  = \frac{1}{2}\int {{d^3}x\,\sum\limits_i {d{\pi ^i}}  \wedge d{\phi _i}}
\end{equation}
Where $\phi _i $ and ${{\pi ^i}}$ are
fields and conjugate momentum fields respectively. By imposing
constraints and gauge fixing conditions on the field expansions,
we obtain them in terms of the physical modes. Using the expansions of the fields in terms of a set of physical modes, normally gives
\begin{equation}
\Omega  = \sum\limits_{i,j} {\int {{d^3}k\,{\omega _{ij}}d{a_i}(k,t) \wedge d{a_j}(k,t)}} \label{two form}
\end{equation}
where ${{a_i}}$ are physical modes and ${{\omega _{ij}}}$ is the
symplectic matrix. Finally by inverting the symplectic
matrix, we get the Dirac brackets of the physical modes
\begin{equation}
{\left\{ {{a_i}(k,t),{a_j}(k',t)} \right\}_{D.B}} = {\omega ^{ij}}{\delta ^3}(k - k')\label{diracbracket}
\end{equation}
where ${\omega ^{ij}}{\omega _{jk}} = \delta _k^i$.

\item[D)] In order to consider the dynamics of the theory we should construct the canonical Hamiltonian of the system. Then we should write it in terms of physical modes derived in step(B). By solving the equations of motion of the physical modes based on the Dirac brackets in Eq.\eqref{diracbracket}, one can write the physical modes in terms of certain quantities at a given time, for example $a_{i}(k,0)$. These quantities are called Schr\"odinger modes. The basic Algebra of Dirac brackets, i.e. Eq.\eqref{diracbracket}, can be written in terms of Schr\"odinger modes.

\item[E)] Finally, using Dirac prescription of quantization, we quantize the theory according to
\begin{equation}
\left\{ {\,\,\,\,,\,\,\,\,} \right\}\,\, \to \,\, - i\left[ {\,\,\,\,,\,\,\,\,} \right].
\end{equation}
Note that by converting Schr\"odinger modes to operators, we will have the expansion of the fields in terms of creation and annihilation operators.
\end{enumerate}

\newpage
\section*{Appendix B}
\subsection*{Dirac Brackets of Vector Fields}

Using  brackets \eqref{EMDB} we are able to calculate the Dirac
brackets of the fields as
 \begin{equation} \left\{ {{A_ + }(\tilde{x},{x^
+ }),{\pi ^ - }(\tilde{y},{x^ + })} \right\} = \frac{i}{2}|{x^ - } - {y^ -
}|\theta ({x^ - } - {y^ - }){\partial _ \bot }{\partial _ \bot
}{\delta ^2}({x^ \bot } - {y^ \bot })
 \end{equation}
 \begin{equation}
\left\{ {{A_ + }(\tilde{x},{x^ + }),{\pi ^i}(\tilde{y},{x^ + })} \right\}  =
\frac{i}{2}|{x^ - } - {y^ - }|\theta ({x^ - } - {y^ - }){\partial
_i}{\partial _i}{\delta ^2}({x^ \bot } - {y^ \bot })
\end{equation}
\begin{equation}
\left\{ {{A_ + }(\tilde{x},{x^ + }),{A_ + }(\tilde{y},{x^ + })} \right\}  =
\frac{1}{2}|{x^ - } - {y^ - }{|^2}\theta ({x^ - } - {y^ -
}){\partial _ \bot }{\partial _ \bot }{\delta ^2}({x^ \bot } - {y^
\bot })
\end{equation}
\begin{equation}
\left\{ {{A_ + }(\tilde{x},{x^ + }),{A_i}(\tilde{y},{x^ + })} \right\}   =
\frac{i}{2}|{x^ - } - {y^ - }|\theta ({x^ - } - {y^ - }){\partial
_i}{\delta ^2}({x^ \bot } - {y^ \bot })
\end{equation}
\begin{equation}
\left\{ {{A_i}(\tilde{x},{x^ + }),{A_j}(\tilde{y},{x^ + })} \right\} =
\frac{i}{2}\delta _i^j\theta ({x^ - } - {y^ - }){\delta ^2}({x^
\bot } - {y^ \bot })
\end{equation}
\begin{equation}
\left\{ {{A_i}(\tilde{x},{x^ + }),{\pi ^ - }(\tilde{y},{x^ + })} \right\}  =
\frac{1}{2}\theta ({x^ - } - {y^ - }){\partial _i}{\delta ^2}({x^
\bot } - {y^ \bot })
\end{equation}
\begin{equation}
\left\{ {{A_i}(\tilde{x},{x^ + }),{\pi ^j}(\tilde{y},{x^ + })} \right\} = \frac{{ - 1}}{2}\delta _i^j\theta ({x^ - } - {y^ - }){\partial _i}{\delta ^2}({x^ \bot } - {y^ \bot })
\end{equation}
\begin{equation}
\left\{ {{\pi ^ - }(\tilde{x},{x^ + }),{\pi ^ - }(\tilde{y},{x^ + })} \right\} = \frac{i}{2}\theta ({x^ - } - {y^ - }){\partial _ \bot }{\partial _ \bot }{\delta ^2}({x^ \bot } - {y^ \bot })
\end{equation}
\begin{equation}
\left\{ {{\pi ^ - }(\tilde{x},{x^ + }),{\pi ^i}(\tilde{y},{x^ + })} \right\} = \frac{{ - i}}{2}\theta ({x^ - } - {y^ - }){\partial _i}{\partial _i}{\delta ^2}({x^ \bot } - {y^ \bot })
\end{equation}
\begin{equation}
\left\{ {{\pi ^i}(\tilde{x},{x^ + }),{\pi ^j}(\tilde{y},{x^ + })} \right\} = \frac{i}{2}\delta _i^j\theta ({x^ - } - {y^ - }){\partial _i}{\partial _i}{\delta ^2}({x^ \bot } - {y^ \bot })
\end{equation}
By transforming $ \left\{ {\,\,\,\,,\,\,\,\,} \right\}\,\, \to
\,\, - i\left[ {\,\,\,\,,\,\,\,\,} \right] $ according to Dirac
prescription of quantization, we finally achieve the quantized electromagnetic
theory in light-cone coordinates.

\end{document}